\documentclass[11pt,showpacs,preprintnumbers,amsmath,amssymb]{article}
\usepackage{graphicx}
\usepackage{amssymb}
\usepackage{amsmath}
\makeatletter \oddsidemargin -0.1in \evensidemargin -.1in
\newcommand{\singlespacing}{\let\CS=\@currsize\renewcommand{\baselinestretch}{1.3}\tiny\CS}
\newcommand{\doublespacing}{\let\CS=\@currsize\renewcommand{\baselinestretch}{1.3}\tiny\CS}
\begin{document}
\begin{center}
\textbf{Modulation instability of obliquely propagating ion acoustic waves in a collisionless magnetized plasma consisting of nonthermal and isothermal electrons}\\
S. Dalui$^{1}$ and A. Bandyopadhyay$^{1,*}$\\
\textit{$^{1}$ Department of Mathematics, Jadavpur University, Kolkata - 700 032, India.}\\
$^{*}$ \textit{e-mail: abandyopadhyay1965@gmail.com}
\end{center}

\noindent \textbf{Abstract -} We have studied the modulation instability of obliquely propagating ion acoustic waves in a collisionless magnetized warm plasma consisting of warm adiabatic ions and two different species of electrons at different temperatures. We have derived a nonlinear Schr{\"o}dinger equation using the standard reductive perturbation method to describe the nonlinear amplitude modulation of ion acoustic wave satisfying the dispersion relation of ion acoustic wave propagating at an arbitrary angle to the direction of the external uniform static magnetic field. We have investigated the correspondence between two nonlinear Schr{\"o}dinger equations $-$ one describes the amplitude modulation of ion acoustic waves propagating along any arbitrary direction to the direction of the magnetic field and other describes the amplitude modulation of ion acoustic waves propagating along the direction of the magnetic field. We have derived the instability condition and the maximum growth rate of instability of the modulated ion acoustic wave. We have seen that the region of existence of maximum growth rate of instability decreases with increasing values of the magnetic field intensity whereas the region of existence of the maximum growth rate of instability increases with increasing $\cos \theta$, where $\theta$ is the angle of propagation of the ion acoustic wave with the external uniform static magnetic field. Again, the maximum growth rate of instability increases with increasing $\cos \theta$ and also this maximum growth rate of instability increases with increasing $\beta_{e}$ upto a critical value of the wave number, where $\beta_{e}$ is the parameter associated with the nonthermal distribution of hotter electron species. 

\section{INTRODUCTION} 
The observations of electric field structures by the Freja Satellite [1] in the auroral zone, Viking Satellite [2, 3] and FAST satellite in the auroral region [4-8], indicate the presence of the fast energetic electrons along with the isothermal electrons. The fast energetic electrons can be modeled by the nonthermal distribution of Cairns et al. [9]. Although there is no general procedure for the formation of the fast energetic electrons in the space plasma. So, the cooler electron species follows Maxwell-Boltzmann distribution and the hotter electron species can be taken as the nonthermal distribution of Cairns et al. [9]. Again, Yu \& Luo [10] reported that for phenomena on long time scales, one can consider electrons into two different species if the electrons are physically separated in space/time domain of interest. So, Maxwell - Boltzmann distributed electrons and Cairns [9] distributed nonthermal electrons can be considered as two different electron species only when those electron species are physically separated in the phase space by external or self - consistent fields. On the basis of this assumption, several authors [11-19] considered two populations of electrons at different temperatures. Several authors [16, 20-23] considered different multi-species plasmas containing two different electron species at different temperatures, where the cooler electron species follows the Maxwell-Boltzmann distribution whereas the hotter electron species is modeled by the nonthermal distribution as prescribed by Cairns et al. [9]. In the present paper, we have investigated modulation instability (MI) of ion acoustic (IA) waves propagating along any arbitrary direction to the direction of the external uniform static magnetic field in a collisionless magnetized electron-ion plasma consisting of warm adiabatic ions and a superposition of two distinct populations of electrons at different temperatures.

Several authors [23-29] investigated MI of IA waves in different magnetized plasma systems. But, in the above mentioned papers, the direction of propagation of the IA wave is taken parallel to the direction of the magnetic field. Murtaza \& Salahuddin [30] studied the MI of obliquely propagating IA waves in a collisionless magnetized electron-ion plasma consisting of isothermal electrons. Jehan et al. [31] have investigated MI of low-frequency electrostatic ion waves in magnetized electron-positron-ion plasma. In both the papers [30, 31], authors have used Krylov-Bogoliubov-Mitropolsky (KBM) method [32] to derive the nonlinear Schr{\"o}dinger equation (NLSE). Recently, Ghosh and Banerjee [33] derived a NLSE using the multiple scale perturbation method to describe the MI of obliquely propagating IA waves in a collisionless magnetized electron-positron-ion plasma.

Recently, Dalui et al. [23] have derived a three dimensional nonlinear Schr{\"o}dinger equation to describe the MI of IA waves propagating along the direction of the magnetic field in a collisionless magnetized warm plasma consisting of nonthermal [9] and isothermal electrons. In the present paper, for the first time, we have investigated MI of IA waves propagating at an arbitrary angle to the direction of the external uniform static magnetic field in the same plasma system of Dalui et al. [23]. So, the present paper is an extension of the earlier paper of Dalui et al. [23] in the following directions:
\begin{enumerate}
	\item Starting from the appropriate basic equations, our aim is to make a systematic development to derive an appropriate nonlinear Schr{\"o}dinger equation (NLSE) which can effectively describe the MI of IA waves propagating along any arbitrary direction with respect to the direction of the external uniform static magnetic field.
	\item Our aim is to make a correspondence between the NLSE describing the amplitude modulation of IA waves propagating along the direction of the magnetic field and the NLSE describing the amplitude modulation of IA waves propagating along any arbitrary direction to the direction of the magnetic field. In fact, we have seen the following points:
\begin{itemize}
	\item The linearized dispersion relation of the IA wave obtained in the present paper is exactly same as that of Dalui et al. [23] if we put $\theta = 0$, where $\theta$ is the angle between the direction of propagation of IA wave and the direction of the external uniform static magnetic field.
	\item The expression of the group velocity of the IA wave obtained in the present paper is exactly same as that of Dalui et al. [23] if we put $\theta = 0$.
	\item The NLSE obtained in the present paper for describing the amplitude modulation of IA waves propagating along any arbitrary direction to the direction of the magnetic field is exactly same as that of Dalui et al. [23] if we put $\theta = 0$ and if we ignore the perpendicular dispersive terms of the NLSE equation of Dalui et al. [23], i.e., if we ignore the weak dependence of the spatial coordinates perpendicular to the direction of the magnetic field. 
\end{itemize}   
\item Using the NLSE, our aim is to find the instability condition of the modulated IA waves propagating along any arbitrary direction.
\item Our aim is to present a through analysis of the instability condition with respect to $n_{2} (= \cos \theta)$. 
\item We want to analyze the characteristics of the growth rate of instability and the maximum growth rate of instability with respect to $n_{2} (= \cos \theta)$.
\item We want to analyze the existence region of $n_{2}$ with respect to $k$ for instability of the modulated IA wave, where $k$ is the wave number of IA wave.    
\end{enumerate}
\section{FORMULATION OF THE PROBLEM}  
In the present paper, we have investigated the MI of obliquely propagating IA waves in a collisionless magnetized plasma consisting of nonthermal and isothermal electrons, whereas Dalui et al. [23] have investigated the MI of IA waves propagating along the direction of the magnetic field in the same plasma system. So, here the IA wave is propagating along a direction making an angle $\theta$ with the external uniform static magnetic field, whereas in Dalui et al. [23], the value of $\theta$ is equal to 0. 

So, we have used the equations (1) - (14) of Dalui et al. [23] and consequently without repeating all those equations, we want to mention the final form of the equation of continuity of ions, equation of motions along $x$, $y$ and $z$ axis, and the Poisson equation of Dalui et al. [23] for easy readability of the present paper.
\begin{eqnarray}\label{P4_continuity_equation}
\frac{d n}{d t} +n\Big( \frac{\partial u}{\partial x}+ \frac{\partial v}{\partial y}+ \frac{\partial w}{\partial z}\Big) = 0 ,
\end{eqnarray}
\begin{eqnarray}\label{P4_Equation_of_motion_x100}
\frac{d u}{d t}  = -\frac{\partial H}{\partial x} + \omega_{c}v ,
\end{eqnarray}
\begin{eqnarray}\label{P4_Equation_of_motion_y100}
\frac{d v}{d t}  = -\frac{\partial H}{\partial y} - \omega_{c}u ,
\end{eqnarray}
\begin{eqnarray}\label{P4_Equation_of_motion_z100}
\frac{d w}{d t}  = -\frac{\partial H}{\partial z} ,
\end{eqnarray}
\begin{eqnarray}\label{P4_Poisson_equation_1}
\nabla^{2}\phi = h_{0} + h_{1} \phi + h_{2} \phi^{2} + h_{3} \phi^{3} - n ,
\end{eqnarray}
where
\begin{eqnarray}\label{P4_Form_of_H}
H = \phi + \frac{\sigma \gamma n^{\gamma-1}}{\gamma-1} ,
\end{eqnarray}
\begin{eqnarray}
\frac{d}{dt} = \frac{\partial }{\partial t}+u\frac{\partial }{\partial x}+v\frac{\partial }{\partial y}+w\frac{\partial }{\partial z} ,
\end{eqnarray}
\begin{eqnarray}
\nabla^{2} = \frac{\partial^{2}}{\partial x^{2}}+\frac{\partial^{2}}{\partial y^{2}}+\frac{\partial^{2}}{\partial z^{2}}.
\end{eqnarray}

Here, $h_{0} = 1$, $h_{1}=(1-\beta_{e} \bar{n}_{c0} \sigma_{c})$, $h_{2} = \frac{1}{2} \big[\bar{n}_{s0} \sigma_{s}^{2} + \bar{n}_{c0} \sigma_{c}^{2} \big]$, $h_{3} = \frac{1}{6} \big[\bar{n}_{s0} \sigma_{s}^{3} + \bar{n}_{c0} (1+3 \beta_{e}) \sigma_{c}^{3}  \big]$ with $\bar{n}_{s0}=\frac{n_{s0}}{n_{0}}$, $\bar{n}_{c0}=\frac{n_{c0}}{n_{0}}$, $\sigma_{c}=\frac{T_{ef}}{T_{ce}}$, $\sigma_{s}=\frac{T_{ef}}{T_{se}}$, and $T_{ef}$ is given by the following equation:
\begin{eqnarray}\label{P4_T_ef}
\frac{ n_{c0} + n_{s0}}{T_{ef}} =  \frac{ n_{c0}}{T_{ce}} + \frac{ n_{s0}}{T_{se}},
\end{eqnarray}
where ${n}_{0}$, ${n}_{c0}$, ${n}_{s0}$, $T_{ce}$ and $T_{se}$ are unperturbed ion number density, unperturbed nonthermal electron number density, unperturbed isothermal electron number density, average temperature of nonthermal electrons and average temperature of isothermal electrons respectively and $\beta_{e}$ is the parameter associated with the nonthermal distribution function of Cairns et al. [9] and the physically admissible bounds of $\beta_{e}$ is given by $0 \leq \beta_{e} \leq \frac{4}{7} \approx 0.6$.

We have used the notations $n$, $n_{ce}$, $n_{se}$, $\omega_{c}$, $\textbf{u}=(u,v,w)$, $\phi$, $(x,y,z)$ and $t$ for the ion number density, the nonthermal electron number density, the isothermal electron number density, the ion cyclotron frequency, the ion fluid velocity vector, the electrostatic potential, the spatial variables  and time, respectively. Here $n$, $n_{ce}$, $n_{se}$, $\omega_{c}$, $\textbf{u}=(u,v,w)$,  $\phi$, $(x,y,z)$ and $t$ are normalized variables and these quantities have been normalized by $n_{0}$, $n_{0}$, $n_{0}$, $\omega_{pi}(= \sqrt{4\pi n_{0} e^{2}/m})$, $c_{s}$ $(= \sqrt{K_{B} T_{ef}/m})$, $K_{B} T_{ef}/e$, $\lambda_{Df}$ $(= \sqrt{K_{B} T_{ef}/4\pi n_{0} e^{2}} )$ and $\omega_{pi}^{-1}$, where $\sigma = T_{i}/T_{ef}$ and $\gamma (=\frac{5}{3})$ is the ratio of two specific heats. Again, $K_{B}$ is the Boltzmann constant, $m$ is the mass of an ion, $-e$ is the charge of an electron, $T_{i}$ is the average ion temperature. Again, the basic parameters of the present plasma system are as follows: $\gamma$, $\sigma$, $\sigma_{sc} = \frac{T_{se}}{T_{ce}}$, $n_{sc} = \frac{n_{s0}}{n_{c0}}$, $\beta_{e}$ and $\omega_{c}$. With respect to the basic parameters $\sigma_{sc}$ and $n_{sc}$, the expressions of $\bar{n}_{s0}=\frac{n_{s0}}{n_{0}}$, $\bar{n}_{c0}=\frac{n_{c0}}{n_{0}}$, $\sigma_{s}$ and $\sigma_{c}$ can be written as follows:
\begin{eqnarray}\label{P4_form_of_sigma_c1}
\Big(\bar{n}_{s0}~,~\bar{n}_{c0}\Big) =  \bigg(\frac{n_{sc}}{1+n_{sc}}~,~\frac{1}{1+n_{sc}}\bigg),
\end{eqnarray}
\begin{eqnarray}\label{P4_form_of_sigma_c2}
\Big(\sigma_{s}~,~\sigma_{c} \Big) = \Big(\frac{1+n_{sc}}{\sigma_{sc}+n_{sc}}~,~\sigma_{sc}\frac{1+n_{sc}}{\sigma_{sc}+n_{sc}}\Big),
\end{eqnarray}
where we have used the charge neutrality condition: $n_{c0} + n_{s0} = n_{0}$, equation (\ref{P4_T_ef}) to find the expressions $\bar{n}_{s0}$, $\bar{n}_{c0}$, $\sigma_{s}$ and $\sigma_{c}$.

Dalui et al. [23] have discussed the MI of IA waves propagating along the direction of the magnetic field ($\textbf{B} = B_{0}\hat{\textbf{z}}$) in a collisionless magnetized plasma. The linear dispersion relation of the IA wave propagating along the direction of the magnetic field ($\textbf{B} = B_{0}\hat{\textbf{z}}$) can be written as
\begin{eqnarray}\label{P4_dispersion_relation_linear_100}
\frac{k_{\parallel}^{2}}{\omega^{2}} = \frac{k_{\parallel}^2+h_{1}}{1+\sigma\gamma(k_{\parallel}^2+h_{1})} ,
\end{eqnarray}
where $\omega$ is the wave frequency and $k_{\parallel}$ is the wave number. This dispersion relation can be easily obtained by linearizing the equations (\ref{P4_continuity_equation}), (\ref{P4_Equation_of_motion_x100}), (\ref{P4_Equation_of_motion_y100}), (\ref{P4_Equation_of_motion_z100}) and (\ref{P4_Poisson_equation_1}) with respect to the perturbed dependent variables, and finally, assuming space-time dependence of the perturbed dependent variables to be of the form $\exp [i(k_{\parallel}z-\omega t)]$. The dispersion relation (\ref{P4_dispersion_relation_linear_100}) is exactly same as the dispersion relation (15) of Dalui et al. [23] with slight modification in notations only. The dispersion relation (\ref{P4_dispersion_relation_linear_100}) or the dispersion relation (15) of Dalui et al. [23] is free from the effect of magnetic field. 

To get the effect of the magnetic field in the linear dispersion relation of the IA wave, it is essential to consider the oblique propagation of IA waves with the external uniform static magnetic field ($\textbf{B} = B_{0}\hat{\textbf{z}}$). Considering the IA waves propagating along a direction having direction cosines $(l_{2},~m_{2},~n_{2})$ and assuming space-time dependence of the perturbed dependent variables to be of the form $\exp [i\{k(l_{2}x+m_{2}y+n_{2}z)-\omega t\}]$, the linear dispersion relation of the IA wave can be written as
\begin{eqnarray}\label{P4_dispersion_relation_linear_200}
\frac{k_{\perp}^{2}}{\omega^{2}-\omega_{c}^{2}} + \frac{k_{\parallel}^{2}}{\omega^{2}} = \frac{k^2+h_{1}}{1+\sigma\gamma(k^2+h_{1})} ,
\end{eqnarray}
where $k_{\perp}^{2} = k^2(l_{2}^{2}+m_{2}^{2})$, $k_{\parallel}^{2} = k^2 n_{2}^{2}$ and $l_{2}^{2}+m_{2}^{2}+n_{2}^{2}=1$.

Here, $\omega$ and $\omega_{c}$ both are normalized by $\omega_{pi}$ and consequently for ion acoustic wave we have the following inequality: $\omega <<\omega_{c}$. Using this inequality ($\omega <<\omega_{c}$), the dispersion relation (\ref{P4_dispersion_relation_linear_200}) can be written as
\begin{eqnarray}\label{P4_dispersion_relation_linear_300}
\omega=k_{\parallel}\Bigg[\frac{k_{\perp}^{2}}{\omega_{c}^{2}} +  \frac{k^2+h_{1}}{1+\sigma\gamma(k^2+h_{1})} \Bigg]^{\displaystyle-\frac{1}{2}},
\end{eqnarray}
or
\begin{equation}\label{P4_dispersion_relation_linear_500}
\frac{\omega}{k}=n_{2}\Bigg[\frac{k^{2}(1-n_{2}^{2})}{\omega_{c}^{2}} +  \frac{k^2+h_{1}}{1+\sigma\gamma(k^2+h_{1})} \Bigg]^{-\frac{1}{2}}. 
\end{equation}

Now, the presence of $\omega_{c}$ in the dispersion relation $\omega=\omega(k)$ as given in the equation  (\ref{P4_dispersion_relation_linear_300}) or the equation (\ref{P4_dispersion_relation_linear_500}) shows that the IA wave is not free from the effect of magnetic field. Therefore, in order to investigate the MI of IA waves propagating along a direction having direction cosines $(l_{2},~m_{2},~n_{2})$ in a collisionless magnetized plasma, we consider the following transformation:
\begin{eqnarray}\label{P4_Transformation}
X = l_{2} x + m_{2} y + n_{2} z .
\end{eqnarray}
Under the above transformation of the spatial variables, the equations (\ref{P4_continuity_equation}) - (\ref{P4_Poisson_equation_1}) assume the following form: 
\begin{eqnarray}\label{P4_Continuity_equation_2}
\frac{\partial n}{\partial t} + \Psi \frac{\partial n}{\partial X} + n\frac{\partial \Psi}{\partial X} = 0  ,
\end{eqnarray}
\begin{eqnarray}\label{P4_x_component_of_equation_of_motion_2}
\frac{\partial u}{\partial t}+ \Psi \frac{\partial u}{\partial X}+ l_{2}\frac{\partial H}{\partial X}-\omega_{c}v=0   , 
\end{eqnarray}
\begin{eqnarray}\label{P4_y_component_of_equation_of_motion_2}
\frac{\partial v}{\partial t}+ \Psi \frac{\partial v}{\partial X}+ m_{2}\frac{\partial H}{\partial X}+\omega_{c}u=0   ,
\end{eqnarray}
\begin{eqnarray}\label{P4_z_component_of_equation_of_motion_2}
\frac{\partial w}{\partial t}+ \Psi \frac{\partial w}{\partial X}+ n_{2}\frac{\partial H}{\partial X}=0   ,
\end{eqnarray}
\begin{eqnarray}\label{P4_Poisson_equation_2}
\frac{\partial^{2} \phi}{\partial X^{2}}=(h_{0}-n)+h_{1}\phi+h_{2}\phi^{2}+h_{3}\phi^{3}   ,
\end{eqnarray}
where
\begin{eqnarray}\label{P4_Form_of_psi}
\Psi = l_{2} u + m_{2} v + n_{2} w    .
\end{eqnarray}

If the direction of propagation of the IA wave makes an angle $\theta$ with the direction of the external uniform static magnetic field, then
\begin{eqnarray}\label{definition_l_2_m_2_n_2}
n_{2} = \cos \theta ~,~ l_{2}^{2} + m_{2}^{2} = \sin^{2} \theta.
\end{eqnarray}
If we put $\theta = 0$, i.e., $n_{2}=1$ ($\Leftrightarrow l_{2}^{2} + m_{2}^{2} = 0 \Leftrightarrow l_{2}= m_{2} = 0$) then the dispersion relation (\ref{P4_dispersion_relation_linear_300}) or (\ref{P4_dispersion_relation_linear_500}) reduces to the dispersion relation (\ref{P4_dispersion_relation_linear_100}), which describes the dispersion relation of IA waves propagating along the direction of the external uniform static magnetic field.

So, the equations (\ref{P4_Continuity_equation_2}) - (\ref{P4_Form_of_psi}) along with the equation (\ref{P4_Form_of_H}) are the basic equations to investigate the MI of IA waves propagating along any arbitrary direction.

\section{DERIVATION OF THE NLSE}
In order to study the MI of IA waves propagating along $X - $axis in a collisionless magnetized plasma, we have used the following stretchings of the spatial coordinate and time:
\begin{eqnarray}\label{P4_Stretching}
\xi=\epsilon (X-V_{g}t) , \tau=\epsilon^{2} t ,
\end{eqnarray}
where $\epsilon$ is a small parameter, $V_{g}$ is a constant and we have used exactly the same method as discussed in section III of the earlier paper of Dalui et al. [23] to find the stretched spatial variables $\xi$ and stretched time $\tau$ as given in the equation (\ref{P4_Stretching}) of the present paper. 

To make a balance between the nonlinear and dispersive terms, we take the following perturbation of the dependent field quantities:
\begin{eqnarray}\label{P4_Perturbation_of_f}
f &=& f^{(0)} + \sum_{l=1}^{\infty} \epsilon^{l} ~ \sum_{a=-\infty}^{\infty} f_{a}^{(l)}(\xi,\tau) \exp{[ia \psi]} ,
\end{eqnarray}
where $\psi=kX-\omega t$, $k$ is the wave number, $\omega$ is the wave frequency, $f$ = $n$, $u$, $v$, $w$, $\phi$  with $n^{(0)}=1$, $u^{(0)}=v^{(0)}=w^{(0)}=0$, $\phi^{(0)}=0$. Here we have also used the terminology: $f_{-a}^{(l)} = \bar{f}_{a}^{(l)}$, where `bar' indicates the conjugate of a complex number.

Following the same analysis of Dalui et al. [23], we can take $\bar{f}_{0}^{(l)}=0$ for any admissible value of $l$ along with the following consistency conditions:
\begin{eqnarray}
&(i)& n_{0}^{(1)}=u_{0}^{(1)}=v_{0}^{(1)}=w_{0}^{(1)}= \phi_{0}^{(1)}=0, \label{P4_consistency_1} \\
&(ii)& n_{a}^{(l)}=0, u_{a}^{(l)}=v_{a}^{(l)}=w_{a}^{(l)}=0, \phi_{a}^{(l)}=0 \mbox{   for   }   l<|a|.  \label{P4_consistency_2}
\end{eqnarray}

Substituting the perturbation expansions for $n$, $u$, $v$, $w$ and $\phi$, according to the law as given in (\ref{P4_Perturbation_of_f}), into the equations (\ref{P4_Continuity_equation_2}) - (\ref{P4_Poisson_equation_2}) and collecting the terms of different powers of $\epsilon $, we get a sequence of equations of different orders. From each equation of a particular order, one can generate another sequence of equations for different harmonics by changing the values of $a$.

\subsection{\label{P4_First Order:} First Order : O$(\epsilon)=1$}
It is simple to check that the zeroth harmonic ($a=0$) equations for the continuity equation of ions, the equations describing the motion of ion fluid and the Poisson equation are identically satisfied due to the consistency condition (\ref{P4_consistency_1}). So, at this order, we shall first of all consider the equations of first harmonic ($ a=1 $). Solving the first harmonic ($ a=1 $) equations of the continuity equation of ions and the equation of motion of ions for the unknowns $ n_{1}^{(1)} $, $ u_{1}^{(1)} $, $ v_{1}^{(1)} $ and $ w_{1}^{(1)} $, we get
\begin{eqnarray}\label{P4_n_11}
n_{1}^{(1)} = \frac{\Lambda}{1-\sigma\gamma\Lambda} \phi_{1}^{(1)}   ,
\end{eqnarray}
\begin{eqnarray}\label{P4_u_11}
u_{1}^{(1)} = \frac{\omega k l_{2} + i \omega_{c}k m_{2}}{\omega^{2}-\omega_{c}^{2}} \frac{1}{1-\sigma\gamma\Lambda} \phi_{1}^{(1)}   ,
\end{eqnarray}
\begin{eqnarray}\label{P4_v_11}
v_{1}^{(1)} = \frac{\omega k m_{2} - i \omega_{c}k l_{2}}{\omega^{2}-\omega_{c}^{2}} \frac{1}{1-\sigma\gamma\Lambda} \phi_{1}^{(1)}  ,
\end{eqnarray}
\begin{eqnarray}\label{P4_w_11}
w_{1}^{(1)} = \frac{k n_{2} }{\omega} \frac{1}{1-\sigma\gamma\Lambda} \phi_{1}^{(1)}    ,
\end{eqnarray}  
where
\begin{eqnarray}\label{P4_form_of_lambda}
\Lambda = \frac{k^{2}(\omega^{2} - n_{2}^{2}\omega_{c}^{2} )}{\omega^{2}(\omega^{2}-\omega_{c}^{2})}.
\end{eqnarray}

From the Poisson equation at the first harmonic ($ a=1 $), we get
\begin{eqnarray}\label{P4_Poisson_equation_1st_order_and_1st_harmonic}
n_{1}^{(1)} = (k^{2}+h_{1}) \phi_{1}^{(1)} .
\end{eqnarray}

Therefore, from equations (\ref{P4_n_11}) and (\ref{P4_Poisson_equation_1st_order_and_1st_harmonic}), we get the following dispersion relation
\begin{eqnarray}\label{P4_dispersion_relation}
\frac{k_{\perp}^{2}}{\omega^{2}-\omega_{c}^{2}} + \frac{k_{\parallel}^{2}}{\omega^{2}} = \frac{k^2+h_{1}}{1+\sigma\gamma(k^2+h_{1})} ,
\end{eqnarray}
where we have used the expession of $\Lambda$ as given in (\ref{P4_form_of_lambda}) to get the equation (\ref{P4_dispersion_relation}). The equation (\ref{P4_dispersion_relation}) is exactly same as the dispersion relation (\ref{P4_dispersion_relation_linear_200}) and this equation will assume the dispersion relation (\ref{P4_dispersion_relation_linear_300}) or equivalently the equation (\ref{P4_dispersion_relation_linear_500}) if we use the condition $\omega <<\omega_{c}$ for the ion acoustic mode.

\subsection{Second Order : O$(\epsilon)=2$}
\subsubsection{First Harmonic ($ a=1 $)}:

Solving the first harmonic ($ a=1 $) equations of the continuity equation of ions and the equations describing the motion of ion fluid for the unknowns $n_{1}^{(2)}$, $u_{1}^{(2)}$, $v_{1}^{(2)}$ and $ w_{1}^{(2)} $, we get
\begin{equation}\label{P4_n_12}
n_{1}^{(2)} = \frac{\Lambda}{1-\sigma\gamma\Lambda} \phi_{1}^{(2)}  +  2ik \Bigg[\frac{V_{g}k\{\omega^4 - n_{2}^{2} \omega_{c}^{2}(2\omega^{2}-\omega_{c}^{2})\}}{\omega^{3}(\omega^{2}-\omega_{c}^{2})^2 (1-\sigma\gamma\Lambda)^2}  - \frac{(\omega^{2}-n_{2}^{2}\omega_{c}^{2})}{\omega^{2}(\omega^{2}-\omega_{c}^{2}) (1-\sigma\gamma\Lambda)^2} \Bigg] \frac{\partial \phi_{1}^{(1)}}{\partial \xi}  ,
\end{equation}
\begin{eqnarray}\label{P4_u_12}
u_{1}^{(2)} = \frac{\omega k l_{2} + i \omega_{c}k m_{2}}{\omega^{2}-\omega_{c}^{2}} [ \phi_{1}^{(2)} + \sigma\gamma n_{1}^{(2)}  ]  + \Bigg[ \frac{V_{g}k\{i(\omega^{2}+\omega_{c}^{2})l_{2}-2\omega \omega_{c} m_{2}\}}{(\omega^{2}-\omega_{c}^{2})^{2}(1-\sigma\gamma\Lambda)} \nonumber \\  - \frac{(i\omega l_{2}-\omega_{c} m_{2})}{(\omega^{2}-\omega_{c}^{2})(1-\sigma\gamma\Lambda)} \Bigg] \frac{\partial \phi_{1}^{(1)}}{\partial \xi}   ,
\end{eqnarray}
\begin{eqnarray}\label{P4_v_12}
v_{1}^{(2)} = \frac{\omega k m_{2} - i \omega_{c}k l_{2}}{\omega^{2}-\omega_{c}^{2}} [ \phi_{1}^{(2)} + \sigma\gamma n_{1}^{(2)}  ]  + \Bigg[ \frac{V_{g}k\{i(\omega^{2}+\omega_{c}^{2})m_{2}+2\omega \omega_{c} l_{2}\} }{(\omega^{2}-\omega_{c}^{2})^{2}(1-\sigma\gamma\Lambda)} \nonumber\\
- \frac{(i\omega m_{2}+\omega_{c} l_{2})}{(\omega^{2}-\omega_{c}^{2})(1-\sigma\gamma\Lambda)} \Bigg] \frac{\partial \phi_{1}^{(1)}}{\partial \xi}   ,
\end{eqnarray}
\begin{eqnarray}\label{P4_w_12}
w_{1}^{(2)} =\frac{k n_{2}}{\omega} [ \phi_{1}^{(2)} + \sigma\gamma n_{1}^{(2)} ]  + \frac{i(V_{g}k-\omega)n_{2}}{\omega^2(1-\sigma\gamma\Lambda)} \frac{\partial \phi_{1}^{(1)}}{\partial \xi}  . 
\end{eqnarray}

From of the Poisson equation the first harmonic ($ a=1 $) equation, we get
\begin{eqnarray}\label{P4_Poisson_equation_2nd_order_and_1st_harmonic}
n_{1}^{(2)} = (k^{2}+h_{1}) \phi_{1}^{(2)} -2ik \frac{\partial \phi_{1}^{(1)}}{\partial \xi} .
\end{eqnarray}

From equations (\ref{P4_n_12}) and (\ref{P4_Poisson_equation_2nd_order_and_1st_harmonic}), we get
\begin{eqnarray}\label{P4_equation_1}
\bigg[ \frac{\Lambda}{1-\sigma\gamma\Lambda} - (k^2+h_{1}) \bigg] \phi_{1}^{(2)} + i A \Bigg[ V_{g} - \frac{\omega(\omega^{2}-\omega_{c}^{2}) (\omega^{2}- n_{2}^{2}\omega_{c}^{2})}{k\{\omega^{4} - n_{2}^{2} \omega_{c}^{2} (2\omega^{2}-\omega_{c}^{2})\} } \nonumber\\ + \frac{\omega^{3}(\omega^{2}-\omega_{c}^{2})^2 (1-\sigma\gamma\Lambda)^{2} }{k\{\omega^{4} - n_{2}^{2} \omega_{c}^{2} (2\omega^{2}-\omega_{c}^{2})\}} \Bigg] \frac{\partial \phi_{1}^{(1)}}{\partial \xi} =0   ,
\end{eqnarray}
where
\begin{eqnarray}\label{P4_A}
A =  \frac{2k^{2}}{\omega^{3}} 
\bigg[ \frac{\omega^{4} - n_{2}^{2}\omega_{c}^{2}(2\omega^{2}-\omega_{c}^{2})}{(\omega^{2}-\omega_{c}^{2})^{2}(1-\sigma\gamma\Lambda)^{2}} \bigg] .
\end{eqnarray}

The first term of the equation (\ref{P4_equation_1}) is equal to zero due to the dispersion relation (\ref{P4_dispersion_relation}) of IA waves, and the second
term of the same equation can be made equal to zero if

\begin{eqnarray}\label{P4_V_g}
V_{g} = \frac{\omega}{k} \Bigg[ \frac{(\omega^{2}-\omega_{c}^{2}) (\omega^{2}- n_{2}^{2}\omega_{c}^{2})}{\{\omega^{4} - n_{2}^{2} \omega_{c}^{2} (2\omega^{2}-\omega_{c}^{2})\} } 
- \frac{\omega^{2}(\omega^{2}-\omega_{c}^{2})^2 (1-\sigma\gamma\Lambda)^{2} }{\omega^{4} - n_{2}^{2} \omega_{c}^{2} (2\omega^{2}-\omega_{c}^{2})} \Bigg]  .
\end{eqnarray}

Again, differentiating the dispersion relation (\ref{P4_dispersion_relation}) with respect to $k$ and using the relation (\ref{P4_form_of_lambda}), we get
\begin{eqnarray}\label{P4_V_g_1}
\frac{\partial \omega}{\partial k} = \frac{\omega}{k} \Bigg[  \frac{(\omega^{2}-\omega_{c}^{2}) (\omega^{2}- n_{2}^{2}\omega_{c}^{2})}{\{\omega^{4} - n_{2}^{2} \omega_{c}^{2} (2\omega^{2}-\omega_{c}^{2})\} }  - \frac{\omega^{2}(\omega^{2}-\omega_{c}^{2})^2 (1-\sigma\gamma\Lambda)^{2} }{\omega^{4} - n_{2}^{2} \omega_{c}^{2} (2\omega^{2}-\omega_{c}^{2})} \Bigg]  .
\end{eqnarray}
Therefore, $V_{g}=\frac{\partial \omega}{\partial k}$ and consequently, equation (\ref{P4_equation_1}) is identically satisfied if $V_{g}$ is the group velocity of the IA wave.

Again, it is simple to check that the expression of $V_{g}$ as given in equation (\ref{P4_V_g}) is exactly same as the expression of $V_{g}$ as given in equation (32) of Dalui et al. [23] if we put $\theta = 0$, i.e., $n_{2}=1$ ($\Leftrightarrow l_{2}^{2} + m_{2}^{2} = 0 \Leftrightarrow l_{2}= m_{2} = 0$).

\subsubsection{Second Harmonic ($ a=2 $)}

Solving the second harmonic ($ a=2 $) equations of the continuity equation of ions, the equations describing the motion of ion fluid and the Poisson equation for the unknowns $ \phi_{2}^{(2)} $, $n_{2}^{(2)}$, $u_{2}^{(2)}$, $v_{2}^{(2)}$ and $ w_{2}^{(2)} $, we get
\begin{eqnarray}\label{P4_Form_of_n_22_u_22_v_22_w_22}
\Big( \phi_{2}^{(2)}, ~ n_{2}^{(2)}, ~ u_{2}^{(2)}, ~ v_{2}^{(2)}, ~ w_{2}^{(2)} \Big) 
= \Big (C_{\phi}, ~ C_{n}, ~ C_{u}, ~ C_{v}, ~ C_{w} \Big) [\phi_{1}^{(1)}]^{2}   ,
\end{eqnarray}
where
\begin{equation}\label{P4_Form_of_C_phi}
C_{\phi} = \frac{2h_{2}(1-\sigma\gamma \chi)(1-\sigma\gamma \Lambda)^{2}-C_{1}\Lambda^{2}}{ 2C_{2} } ,
\end{equation}
\begin{eqnarray}\label{P4_Form_of_C_n}
C_{n} = (4k^{2}+h_{1})C_{\phi} + h_{2}  ,
\end{eqnarray}
\begin{eqnarray}\label{P4_Form_of_C_u_2}
C_{u} =  \omega l_{2} C_{c} +  i \omega_{c} m_{2} C_{s} , 
\end{eqnarray}
\begin{eqnarray}\label{P4_Form_of_C_v_2}
C_{v} =  \omega m_{2} C_{c} -  i \omega_{c} l_{2} C_{s} , 
\end{eqnarray}
\begin{equation}\label{P4_Form_of_C_w}
C_{w} = \frac{kn_{2}}{\omega} \Big[ C_{\phi}+\sigma\gamma C_{n}+\frac{\Lambda+\sigma\gamma (\gamma-2)  \Lambda^{2}}{2(1-\sigma\gamma \Lambda)^{2}} \Big]  ,
\end{equation}
\begin{eqnarray}\label{P4_Form_of_C_1}
C_{1} = 2 + \sigma\gamma (\gamma-2) \chi 
+ \frac{ 2\omega^{2}(2\omega^{2}+\omega_{c}^{2}) -n_{2}^{2}\omega_{c}^{2}(7\omega^{2}-\omega_{c}^{2})}{(4\omega^{2}-\omega_{c}^{2})(\omega^{2}-n_{2}^{2}\omega_{c}^{2})} ,
\end{eqnarray}
\begin{equation}\label{P4_Form_of_C_2}
C_{2} = (1-\sigma\gamma \Lambda)^{2} \{\chi - (1-\sigma\gamma \chi)(4k^{2}+h_{1})\} ,
\end{equation}
\begin{eqnarray}\label{P4_Form_of_C_u_c}
C_{c} =  \frac{k}{4\omega^{2}-\omega_{c}^{2}} \Bigg[ \frac{\Lambda (2\omega^{2}+\omega_{c}^{2}) }{(\omega^{2}-\omega_{c}^{2})(1-\sigma\gamma \Lambda)^{2}}   + \frac{2\sigma\gamma (\gamma-2)\Lambda^2}{(1-\sigma\gamma \Lambda)^{2}} + 4(C_{\phi}+\sigma\gamma C_{n})  \Bigg] , 
\end{eqnarray}
\begin{eqnarray}\label{P4_Form_of_C_u_s}
C_{s} =  \frac{k}{4\omega^{2}-\omega_{c}^{2}} \Bigg[ \frac{3\omega^{2}\Lambda }{(\omega^{2}-\omega_{c}^{2})(1-\sigma\gamma \Lambda)^{2}} + \frac{\sigma\gamma (\gamma-2)\Lambda^2}{(1-\sigma\gamma \Lambda)^{2}} + 2(C_{\phi}+\sigma\gamma C_{n}) \Bigg] , 
\end{eqnarray}
\begin{eqnarray}\label{P4_form_of_chi}
\chi = \frac{k^{2}(4\omega^{2} - n_{2}^{2}\omega_{c}^{2} )}{\omega^{2}(4\omega^{2}-\omega_{c}^{2})} .
\end{eqnarray}

\subsubsection{Zeroth Harmonic ($ a=0 $)}
Solving the zeroth harmonic ($ a=0 $) equations of the continuity equation of ions, the equations describing the motion of ion fluid and the Poisson equation for the unknowns $ \phi_{0}^{(2)} $, $n_{0}^{(2)}$, $u_{0}^{(2)}$, $v_{0}^{(2)}$ and $ w_{0}^{(2)} $, we get
\begin{eqnarray}\label{P4_Form_of_n_02_u_02_v_02_w_02}
\Big( \phi_{0}^{(2)}, ~ n_{0}^{(2)}, ~ u_{0}^{(2)}, ~ v_{0}^{(2)}, ~ w_{0}^{(2)} \Big) 
= \Big( D_{\phi}, ~ D_{n}, ~ D_{u}, ~ D_{v}, ~ D_{w} \Big) |\phi_{1}^{(1)}|^{2} ,
\end{eqnarray}
where 
\begin{eqnarray}\label{P4_Form_of_D_phi}
D_{\phi} = \frac{D_{1}}{D_{2}} ,
\end{eqnarray}
\begin{eqnarray}\label{P4_Form_of_D_n}
D_{n} = h_{1} D_{\phi} + 2h_{2} ,
\end{eqnarray}
\begin{eqnarray}\label{P4_Form_of_D_u}
D_{u} = \frac{-2\Lambda \omega k l_{2}}{(\omega^{2}-\omega_{c}^{2})(1-\sigma\gamma\Lambda)^2} , 
\end{eqnarray}
\begin{eqnarray}\label{P4_Form_of_D_v}
D_{v} = \frac{-2\Lambda \omega k m_{2}}{(\omega^{2}-\omega_{c}^{2})(1-\sigma\gamma\Lambda)^2} , 
\end{eqnarray}
\begin{eqnarray}\label{P4_Form_of_D_w}
D_{w} = \frac{n_{2}}{V_{g}} \Big[ D_{\phi} + \sigma\gamma D_{n} - 2k^{2} + \frac{\big(\sigma\gamma(\gamma-2)\Lambda +3 - \frac{2V_{g}k}{\omega} \big)\Lambda}{(1-\sigma\gamma \Lambda)^2} \Big] ,
\end{eqnarray}
\begin{eqnarray}\label{P4_Form_of_D_1}
D_{1} = \frac{ \{ \sigma\gamma (\gamma-2) \Lambda + 3 \} \Lambda n_{2}^{2}}{(1-\sigma\gamma\Lambda)^2} - 2k^2 n_{2}^2 - 2h_{2} (V_{g}^{2}-\sigma\gamma n_{2}^{2}) , 
\end{eqnarray}
\begin{eqnarray}\label{P4_Form_of_D_2}
D_{2} = h_{1}(V_{g}^{2}-\sigma\gamma n_{2}^{2}) - n_{2}^{2} .
\end{eqnarray}

\subsection{Third Order (O($\epsilon$)=3) : First Harmonic ($ a=1 $)}

Solving the continuity equation  of ions and the equations describing the motion of ion fluid for the unknowns $ n_{1}^{(3)} $, $ u_{1}^{(3)} $, $ v_{1}^{(3)} $ and $ w_{1}^{(3)} $, we can express $ n_{1}^{(3)} $, $ u_{1}^{(3)} $, $ v_{1}^{(3)} $ and $ w_{1}^{(3)} $ as a function of $\phi_{1}^{(1)}$, $\phi_{1}^{(2)}$ and $\phi_{1}^{(3)}$ along with their different derivatives with respect to $\xi$ and $\tau$. In particular, $ n_{1}^{(3)} $ can be expressed as
\begin{eqnarray}\label{P4_n_31_1}
n_{1}^{(3)} = \frac{\Lambda}{1-\sigma\gamma\Lambda} \phi_{1}^{(3)} 
- i A  \frac{\partial \phi_{1}^{(1)}}{\partial \tau} + i A  \bigg[ V_{g} - \frac{\omega\{ (\omega^{2}-\omega_{c}^{2}) (\omega^{2}-n_{2}^{2}\omega_{c}^{2}) \}}{k\{\omega^{4} - n_{2}^{2}\omega_{c}^{2}(2\omega^{2}-\omega_{c}^{2})\}} \bigg] \frac{\partial \phi_{1}^{(2)}}{\partial \xi} \nonumber \\ + Q_{1} |\phi_{1}^{(1)}|^{2} \phi_{1}^{(1)} - \bigg[ \frac{P_{1}P_{2} + P_{3} + P_{4} + P_{5}}{1-\sigma\gamma\Lambda} \bigg] \frac{\partial^{2} \phi_{1}^{(1)}}{\partial \xi^{2}},
\end{eqnarray}
where
\begin{eqnarray}\label{P4_P_1}
P_{1} =  \frac{V_{g}k\omega}{\omega^{2}-\omega_{c}^{2}} + \frac{V_{g}k-\omega}{\omega} - \sigma\gamma \Lambda  + \frac{\sigma\gamma V_{g}k^{3}\omega_{c}^{2} \{ \omega^{2} - n_{2}^{2}(2\omega^{2}-\omega_{c}^{2}) \} }{\omega^{3}(\omega^{2}-\omega_{c}^{2})^{2}} ,
\end{eqnarray}
\begin{eqnarray}\label{P4_P_2}
P_{2} =  \frac{2V_{g}k \{\omega^{4} - n_{2}^{2}\omega_{c}^{2}(2\omega^{2}-\omega_{c}^{2}) \} }{\omega^{3}(\omega^{2}-\omega_{c}^{2})^{2}(1-\sigma\gamma\Lambda)^{2}}  - \frac{2(\omega^{2}-n_{2}^{2}\omega_{c}^{2}) }{\omega^{2}(\omega^{2}-\omega_{c}^{2})(1-\sigma\gamma\Lambda)^{2}}    ,
\end{eqnarray}
\begin{eqnarray}\label{P4_P_3}
P_{3} = \frac{ V_{g}^{2}k^{2}\omega_{c}^{2}\{2\omega^{4} - n_{2}^{2}(3\omega^{4}-2\omega^{2}\omega_{c}^{2} + \omega_{c}^{4}) \} }{\omega^{4}(\omega^{2}-\omega_{c}^{2})^{3}(1-\sigma\gamma\Lambda)} - \frac{V_{g}k \omega_{c}^2 \{\omega^{2} -n_{2}^{2}(2\omega^{2}-\omega_{c}^{2})\} }{\omega^{3}(\omega^{2}-\omega_{c}^{2})^{2}(1-\sigma\gamma\Lambda)},
\end{eqnarray}
\begin{eqnarray}\label{P4_P_4}
P_{4} =  \frac{(V_{g}k - \omega)(\omega^{2}-n_{2}^{2}\omega_{c}^{2})}{\omega^{3}(\omega^{2}-\omega_{c}^{2})(1-\sigma\gamma\Lambda)}     ,
\end{eqnarray}
\begin{eqnarray}\label{P4_P_5}
P_{5} = -  \frac{V_{g}k(V_{g}k - \omega)(\omega^{2}-n_{2}^{2}\omega_{c}^{2})}{\omega^{2}(\omega^{2}-\omega_{c}^{2})^{2}(1-\sigma\gamma\Lambda)}     ,
\end{eqnarray}
\begin{eqnarray}\label{P4_Q_1}
Q_{1} = \frac{\Lambda}{(1-\sigma\gamma\Lambda)^2} \Bigg[ \frac{\sigma\gamma(\gamma-2)(\gamma-3)\Lambda^{3}}{2(1-\sigma\gamma\Lambda)^2}  + \{1+\sigma\gamma(\gamma-2)\Lambda\}(C_{n}+D_{n}) + \frac{2 k n_{2} C_{w}}{\omega} \nonumber \\    
+ \frac{2k\big\{(\omega^2 C_{c} + \omega_{c}^{2} C_{s}) (l_{2}^2+m_{2}^2) \big\}}{(\omega^2-\omega_{c}^2)} + \Bigg\{ \frac{ 2k \{\omega^4 - (2\omega^2-\omega_{c}^2) n_{2}^2 \omega_{c}^{2} \} }{\omega (\omega^2-\omega_{c}^2)(\omega^2-n_{2}^{2}\omega_{c}^2) } \nonumber \\
\times (l_{2} D_{u}+m_{2} D_{v}+n_{2} D_{w}) \Bigg\} \Bigg] .
\end{eqnarray}
Again, from the Poisson equation, we get
\begin{eqnarray}\label{P4_Poisson_equation_n_31_final}
n_{1}^{(3)} =  (k^{2}+h_{1}) \phi_{1}^{(3)} - 2ik \frac{\partial \phi_{1}^{(2)}}{\partial \xi}
- \frac{\partial^{2} \phi_{1}^{(1)}}{\partial \xi^{2}}  + \big[2h_{2} (C_{\phi}+D_{\phi}) + 3h_{3} \big]  |\phi_{1}^{(1)}|^{2} \phi_{1}^{(1)}  .
\end{eqnarray}

Now, eliminating $n_{1}^{(3)}$ from the equations (\ref{P4_n_31_1}) and (\ref{P4_Poisson_equation_n_31_final}), we get
\begin{eqnarray}\label{P4_NLSE}
i \frac{\partial \phi_{1}^{(1)}}{\partial \tau}  + P \frac{\partial^{2} \phi_{1}^{(1)}}{\partial \xi^{2}} + Q |\phi_{1}^{(1)}|^{2} \phi_{1}^{(1)} =0 ,
\end{eqnarray}
where
\begin{eqnarray}\label{P4_P}
P = - \frac{1}{A}\Big[ 1 - \frac{P_{1}P_{2} + P_{3} + P_{4}+P_{5}}{1-\sigma\gamma\Lambda}  \Big]  ,
\end{eqnarray}
\begin{eqnarray}\label{P4_Q}
Q = - \frac{1}{A} \Big[ Q_{1} -3h_{3}- 2h_{2}(C_{\phi}+D_{\phi}) \Big]  .
\end{eqnarray}
Here we have removed the terms $\phi_{1}^{(3)}$ and $\frac{\partial \phi_{1}^{(2)}}{\partial \xi}$ using the dispersion relation (\ref{P4_dispersion_relation}) and the compatibility condition (\ref{P4_V_g}) respectively to simplify the equation (\ref{P4_NLSE}).

From the expressions of $P$ and $Q$, it is simple to check that $P$ and $Q$ are the functions of $\gamma$, $\sigma$, $\beta_{e}$, $n_{sc}$, $\sigma_{sc}$, $\omega_{c}$ and $n_{2}$ only, where we have used the relation $l_{2}^{2}+m_{2}^{2}=1-n_{2}^{2}$ to remove $l_{2}^{2}+m_{2}^{2}$ from the resulting expressions of $P$ and $Q$.

It is simple to check that the expressions of $P$ and $Q$ as given in equations (\ref{P4_P}) and (\ref{P4_Q}) are, respectively, same as the expressions of $P$ and $Q$ as given in equations (45) and (46) of Dalui et al. [23] if we put $\theta = 0$, i.e., $n_{2}=1$ ($\Leftrightarrow l_{2}^{2} + m_{2}^{2} = 0 \Leftrightarrow l_{2}= m_{2} = 0$). Therefore, the NLSE (\ref{P4_NLSE}) of the present paper is exactly same as the NLSE (44) of Dalui et al. [23] if we remove the fourth term of equation (44) of Dalui et al. [23]. In fact, the fourth term of equation (44) of Dalui et al. [23] is responsible for the weak dependence of the spatial coordinates perpendicular to the direction of propagation of the wave as described by the first and second equations of (18) of Dalui et al. [23].

\section{\label{P4_sec:Modulation_instability}MODULATIONAL INSTABILITY}
\begin{figure}[ht]
	\begin{center}
		\includegraphics{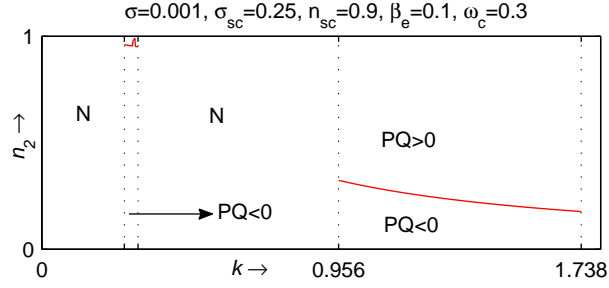}
		\caption{\label{P4_PQ_eq_zero_k_vs_n1} $n_{2}$ is plotted against $k$ when $PQ=0$. This figure shows the regions in $k-n_{2}$ plane described by the inequalities $PQ<0$ and $PQ>0$.}
	\end{center}
\end{figure}
\begin{figure}[ht]
	\begin{center}
		\includegraphics{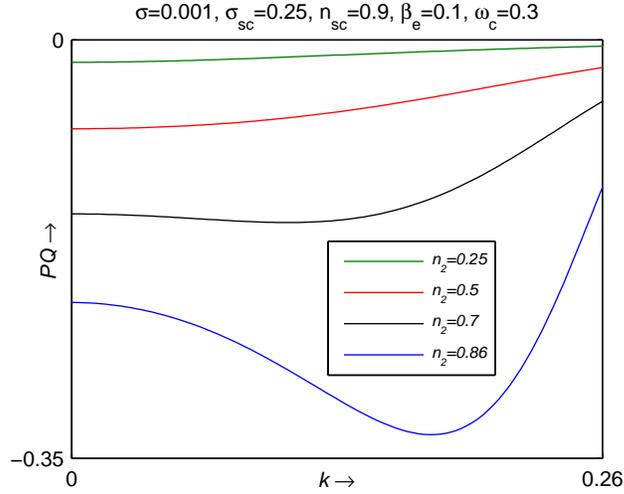}
		\caption{\label{P4_k_vs_PQ_diff_n_1} $PQ$ is plotted against $k$ for different values of $n_{2}$.}
	\end{center}
\end{figure}

\begin{figure}[ht]
	\begin{center}
		\includegraphics{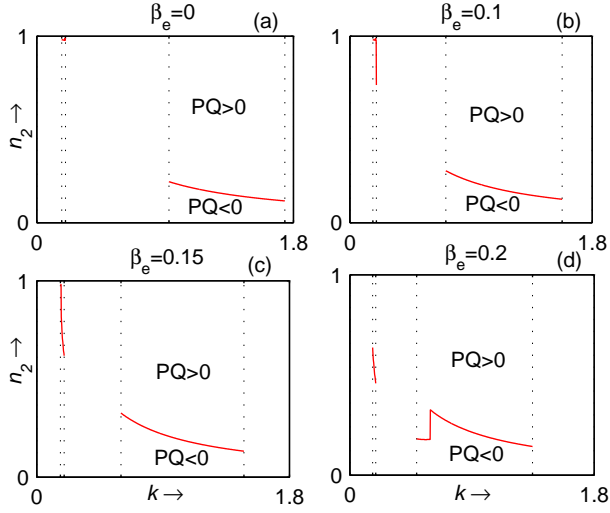}
		\caption{\label{P4_PQ_eq_zero_k_vs_n1_diff_beta_e_subplot} $n_{2}$ is plotted against $k$ when $PQ=0$ for different values of $\beta_{e}$ with $\gamma=5/3$, $\sigma=0.001$, $\sigma_{sc}=0.25$, $n_{sc}=0.25$ and $\omega_{c}=0.2$. These figures show the regions in $k-n_{2}$ plane described by the inequalities $PQ<0$ and $PQ>0$ for different values of $\beta_{e}$. For $PQ<0$, the modulated IA wave is stable whereas for $PQ>0$, the modulated IA wave is stable or unstable according to whether $K\geq K_{c}$ or $K< K_{c}$. }
	\end{center}
\end{figure}
\begin{figure}[ht]
	\begin{center}
		\includegraphics{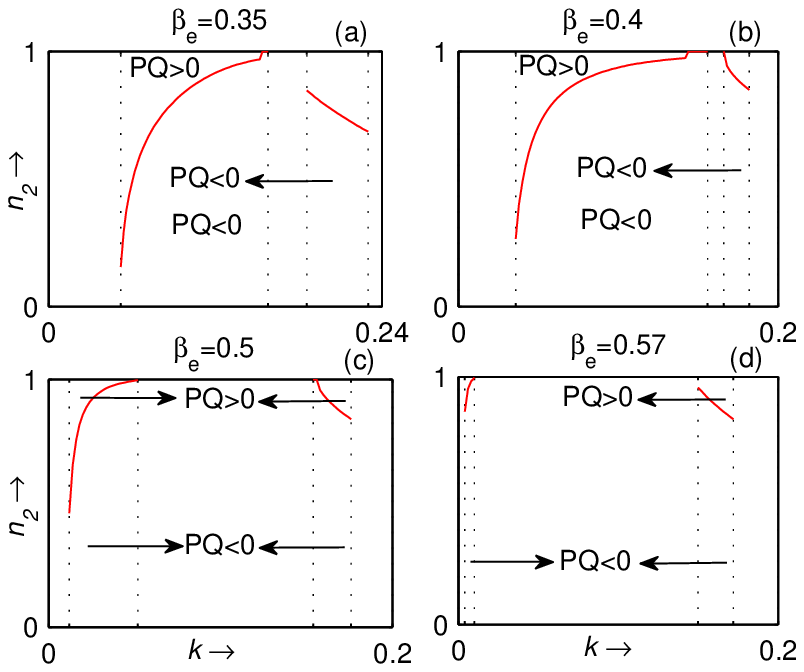}
		\caption{\label{P4_PQ_eq_zero_k_vs_n1_diff_beta_e_subplot_111} $n_{2}$ is plotted against $k$ when $PQ=0$ for different values of $\beta_{e}$ with $\gamma=5/3$, $\sigma=0.001$, $\sigma_{sc}=0.25$, $n_{sc}=0.25$ and $\omega_{c}=0.2$. These figures show the regions in $k-n_{2}$ plane described by the inequalities $PQ<0$ and $PQ>0$ for different values of $\beta_{e}$.}
	\end{center}
\end{figure}


\begin{figure}[ht]
	\begin{center}
		\includegraphics{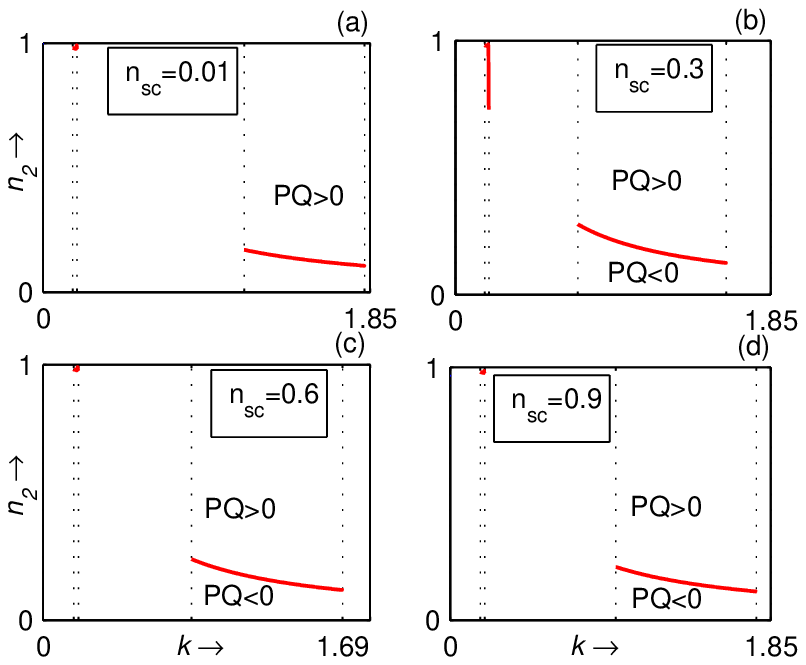}
		\caption{\label{P4_PQ_eq_zero_k_vs_n1_diff_n_sc} $n_{2}$ is plotted against $k$ when $PQ=0$ for different values of $n_{sc}$ with $\gamma=5/3$, $\sigma=0.001$, $\sigma_{sc}=0.25$ and $\beta_{e}=0.1$. These figures show the regions in $k-n_{2}$ plane described by the inequalities $PQ<0$ and $PQ>0$ for different values of $n_{sc}$.}
	\end{center}
\end{figure}

\begin{figure}[ht]
	\begin{center}
		\includegraphics{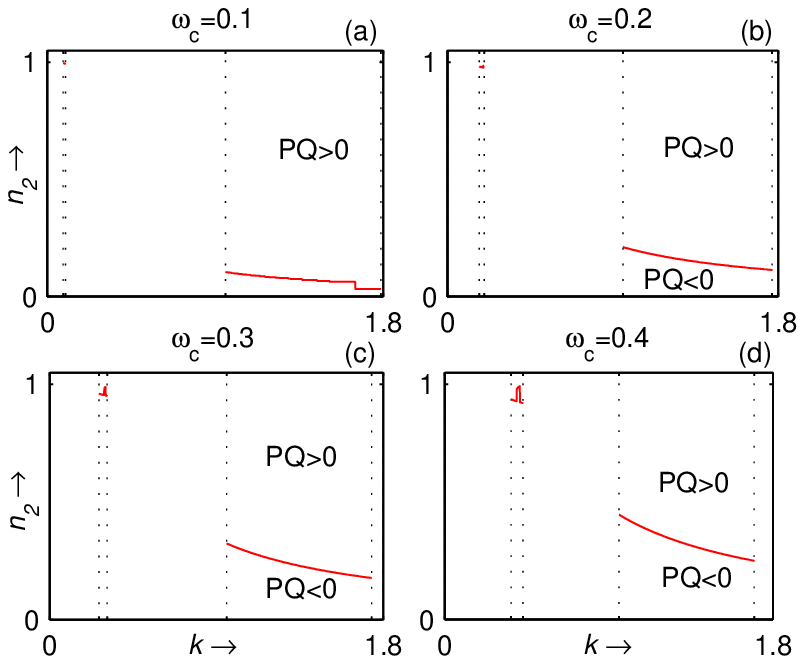}
		\caption{\label{P4_PQ_eq_zero_k_vs_n1_diff_omegha_c_subplot} $n_{2}$ is plotted against $k$ when $PQ=0$ for different values of $\omega_{c}$ with $\gamma=5/3$, $\sigma=0.001$, $\beta_{e}=0.1$, $n_{sc}=0.9$ and $\sigma_{sc}=0.25$. These figures show the regions in $k-n_{2}$ plane described by the inequalities $PQ<0$ and $PQ>0$ for different values of $\omega_{c}$.}
	\end{center}
\end{figure}

\begin{figure}[ht]
	\begin{center}
		\includegraphics[height=8cm]{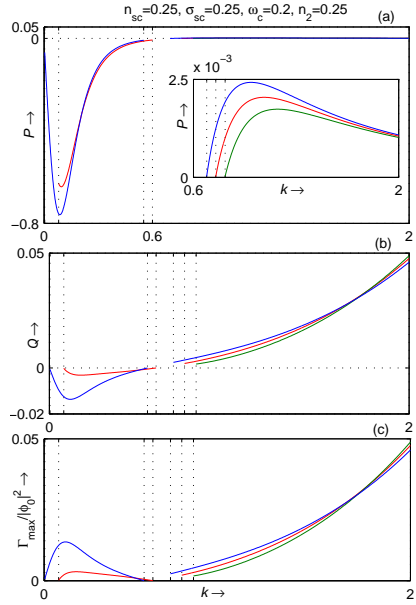}
		\caption{\label{P4_Gamma_max_diff_beta_e} $P$, $Q$ and $\Gamma_{max}/|\phi_{0}|^{2}$ are plotted in (a), (b) and (c) respectively, against $k$ for different values of $\beta_{e}$. Green, red and blue curves of each figure correspond to $\beta_{e}=0$, $\beta_{e}=0.3$ and $\beta_{e}=0.57$ respectively.}
	\end{center}
\end{figure}


\begin{figure}[ht]
	\begin{center}
		\includegraphics[height=8cm]{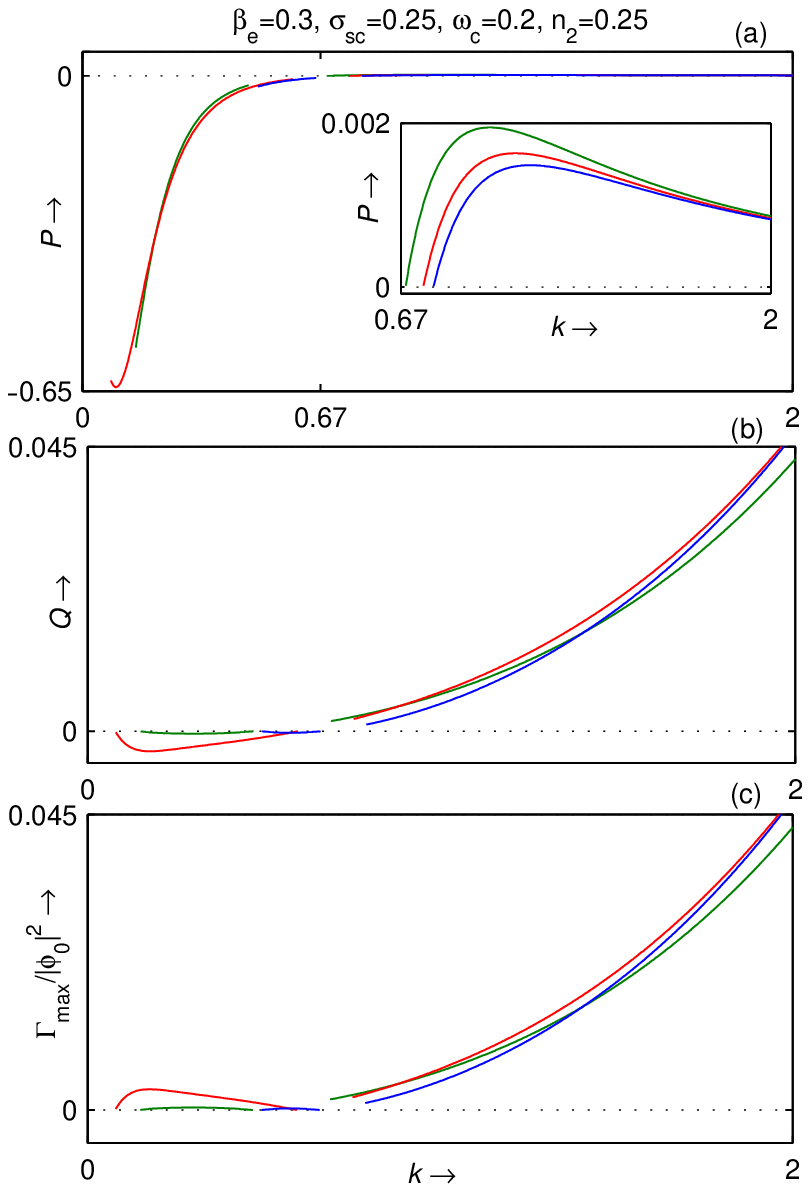}
		\caption{\label{P4_Gamma_max_diff_n_sc} $P$, $Q$ and $\Gamma_{max}/|\phi_{0}|^{2}$ are plotted in (a), (b) and (c) respectively, against $k$ for different values of $n_{sc}$. Green, red and blue curves of each figure correspond to $n_{sc}=0.01$, $n_{sc}=0.25$ and $n_{sc}=0.9$ respectively.}
	\end{center}
\end{figure}
\begin{figure}[ht]
	\begin{center}
		\includegraphics[height=8cm]{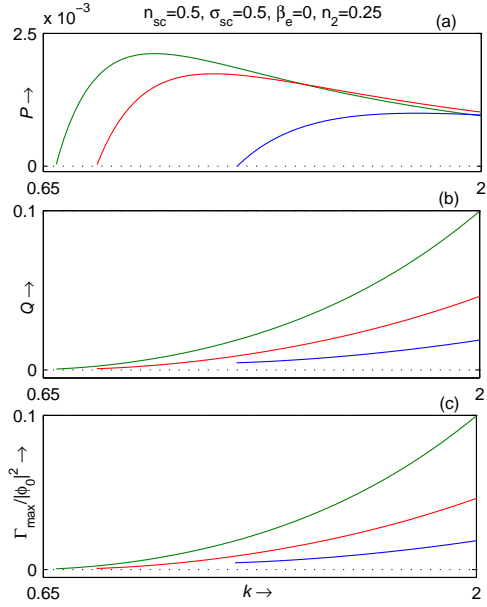}
		\caption{\label{P4_Gamma_max_diff_omega_c} $P$, $Q$ and $\Gamma_{max}/|\phi_{0}|^{2}$ are plotted in (a), (b) and (c) respectively, against $k$ for different values of $\omega_{c}$. Green, red and blue curves of each figure correspond to $\omega_{c}=0.17$, $\omega_{c}=0.2$ and $\omega_{c}=0.3$ respectively.}
	\end{center}
\end{figure}
\begin{figure}[ht]
	\begin{center}
		\includegraphics[height=8cm]{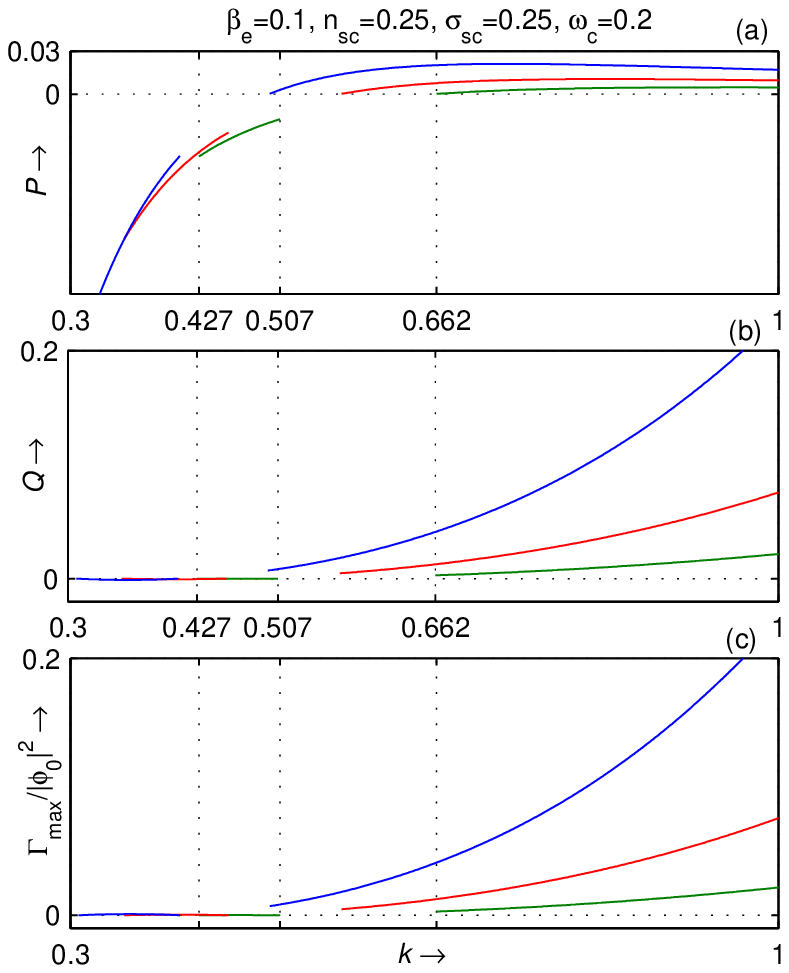}
		\caption{\label{P4_Gamma_max_diff_n_1} $P$, $Q$ and $\Gamma_{max}/|\phi_{0}|^{2}$ are plotted in (a), (b) and (c) respectively, against $k$ for different values of $n_{2}$. Green, red and blue curves of each figure correspond to $n_{2}=0.3$, $n_{2}=0.35$ and $n_{2}=0.4$ respectively.}
	\end{center}
\end{figure}
Following the same method as discussed in section IV of the paper of Dalui et al. [22, 23], we have the following expression of $\Omega^{2}$:
\begin{eqnarray}\label{P4_modulated_dispersion_relation}
\Omega^{2} = [P K^{2}]^{2} \Big( 1 - \frac{2 Q |\phi_{0}|^{2}}{P K^{2}} \Big),
\end{eqnarray}
where $\Omega$ and $K$ are, respectively, the wave frequency and wave number of the modulated IA wave.

From this expression of $\Omega^{2}$, we can conclude the following points regarding the stability of the modulated IA waves:
\begin{itemize}
	\item If $PQ<0$, then $\Omega^{2}$ is strictly positive and consequently the IA wave is always modulationally stable.
	\item If $PQ>0$, then $\Omega^{2} \geq 0$ or $\Omega^{2}<0$ according to whether $K \geq K_{c}$ or $K < K_{c}$, where $K_{c}=\sqrt{\frac{2Q|\phi_{0}|^{2}}{P}}$. Therefore, if $PQ>0$, the IA wave is modulationally stable or unstable according to whether $K \geq K_{c}$ or $K < K_{c}$.
	\item If the IA wave is modulationally unstable, the growth rate of instability $\Gamma$ ($=Im(\Omega)$) is given by the following equation:
	\begin{eqnarray}\label{P4_growth_rate_of_instability}
	\Gamma^{2} = [P K^{2}]^{2} \Big( \frac{2 Q |\phi_{0}|^{2}}{P K^{2}}-1 \Big).
	\end{eqnarray}
	\item If the IA wave is modulationally unstable, $\Gamma$ attains its maximum value $\Gamma_{max}$ at $ K= \frac{K_{c}}{\sqrt{2}} =\sqrt{\frac{Q|\phi_{0}|^{2}}{P}}$, where $\Gamma_{max} = |Q||\phi_{0}|^{2}$.
	\item Although, $\Omega^{2} = [P K^{2}]^{2}$ for $Q=0$ shows that the modulated IA wave is stable but the analysis is erroneous because for $Q=0$ and it is not possible to study the MI of IA waves with the help of the present NLSE. For $Q=0$, the equation (\ref{P4_NLSE}) loses its nonlinearity and consequently, a modified NLSE is necessary to study the MI of IA waves.    
\end{itemize}

\section{SUMMARY AND DISCUSSIONS} 
For obliquely propagating IA waves in a magnetized plasma, we have $\omega<<\omega_{c}$, and consequently the dispersion relation of IA wave is given by the equation (\ref{P4_dispersion_relation_linear_500}). Using the equation (\ref{P4_dispersion_relation_linear_500}), it is simple to check that for any set of values of the parameters involved in the system, the phase velocity ($\frac{\omega}{k}$) and the group velocity ($V_{g}$) are decreasing functions of $k$. Now, as it is observed that the phase velocity and the group velocity of the modulated IA wave takes a very small numerical value at $k = 2$, the value of $k$ is restricted by the inequality $0 < k < 2$.

The MI of obliquely propagating IA waves depends on the coefficients of dispersive and nonlinear terms of the present NLSE (\ref{P4_NLSE}) and we have the following three conditions regarding the stability of the modulated IA waves: 
(i) modulated IA wave is stable if  $PQ<0$,
(ii) modulated IA wave is stable if $K \geq K_{c}$ whenever $PQ>0$ and
(iii) modulated IA wave is unstable if $K < K_{c}$ whenever $PQ>0$.

It is easy to check that $PQ$ is a function of $\sigma$, $\gamma$, $k$, $\beta_{e}$, $n_{sc}$, $\sigma_{sc}$, $\omega_{c}$ and $n_{2}$. Therefore, $PQ$ can be taken as a function of $k$ and $n_{2}$ for fixed values of the other parameters $\sigma$, $\gamma$, $\beta_{e}$, $n_{sc}$, $\sigma_{sc}$ and $\omega_{c}$. Consequently, $PQ=0$ gives a functional relationship between $k$ and $n_{2}$. This functional relation between $k$ and $n_{2}$ is plotted in figure \ref{P4_PQ_eq_zero_k_vs_n1} for fixed values of the other parameters. From figure \ref{P4_PQ_eq_zero_k_vs_n1}, we see that within the intervals $ 0 \leq k \leq 0.266$ and $ 0.31 \leq k \leq 0.956$, there is no functional relation between $k$ and $n_{2}$ for which $PQ=0$. Here, we have used the terminology $N$ to indicate that we are unable to determine the functional relationship between $k$ and $n_{2}$ for which $PQ=0$ for the entire indicated rectangular regions enclosed by $ 0 \leq k \leq 0.266$ and $ 0 \leq n_{2} \leq 1$, and $ 0.31 \leq k \leq 0.956$ and $ 0 \leq n_{2} \leq 1$. In particular, in the rectangular region ($ 0 \leq k \leq 0.266$, $0 \leq n_{2} \leq 1$), the function $PQ$ is strictly negative, i.e., $PQ<0$. Therefore, within that region, there is no functional relation between $k$ and $n_{2}$ for which $PQ=0$. To confirm our analysis, we draw figure \ref{P4_k_vs_PQ_diff_n_1} to explain the behaviour of $PQ$ within the rectangular region of $k-n_{2}$ plane defined by $ 0 \leq k \leq 0.266$ and $0 \leq n_{2} \leq 1$. In figure \ref{P4_k_vs_PQ_diff_n_1}, $PQ$ is plotted against $k$ for different values of $n_{2}$ with $\sigma=0.001$, $\gamma=5/3$, $\beta_{e}=0.1$, $n_{sc}=0.9$, $\sigma_{sc}=0.25$ and $\omega_{c}=0.3$. From figure \ref{P4_k_vs_PQ_diff_n_1}, we see that $PQ < 0$ for all $ 0 \leq k \leq 0.266$ with $n_{2}=0.25$, $n_{2}=0.5$, $n_{2}=0.7$ and $n_{2}=0.86$. So, in the interval $ 0 \leq k \leq 0.266$, $PQ<0$ for any physically admissible value of $n_{2}$ and for this reason there is no functional relationship between $k$ and $n_{2}$ when $PQ=0$ within the interval $ 0 \leq k \leq 0.266$. Finally, from figures \ref{P4_PQ_eq_zero_k_vs_n1} and \ref{P4_k_vs_PQ_diff_n_1}, we see that there exists a region (an interval or union of more than one interval) of $k$ such that $PQ<0$ for any set of given values of the parameters and consequently, MI of the obliquely propagating IA waves are stable for any value of $k$ lying in that region of $k$.

In figures \ref{P4_PQ_eq_zero_k_vs_n1_diff_beta_e_subplot} and \ref{P4_PQ_eq_zero_k_vs_n1_diff_beta_e_subplot_111}, $n_{2}$ is plotted against $k$ when $PQ=0$ for different values of $\beta_{e}$ with $\gamma=5/3$, $\sigma=0.001$, $\sigma_{sc}=0.25$, $n_{sc}=0.25$ and $\omega_{c}=0.2$. In fact, in figure \ref{P4_PQ_eq_zero_k_vs_n1_diff_beta_e_subplot}, $n_{2}$ is plotted against $k$ when $PQ=0$ for (a) $\beta_{e}=0$, (b) $\beta_{e}=0.1$, (c) $\beta_{e}=0.15$ and (d) $\beta_{e}=0.2$ whereas in figure \ref{P4_PQ_eq_zero_k_vs_n1_diff_beta_e_subplot_111}, $n_{2}$ is plotted against $k$ when $PQ=0$ for (a) $\beta_{e}=0.35$, (b) $\beta_{e}=0.4$, (c) $\beta_{e}=0.5$ and (d) $\beta_{e}=0.57$. From figure \ref{P4_PQ_eq_zero_k_vs_n1_diff_beta_e_subplot}, we see that the interval of $k$ for the existence of the curve described by the relation $PQ=0$ in $k-n_{2}$ plane increases with increasing $\beta_{e}$ and it is simple to check that there exists a critical value $\beta_{e}^{(c)}= 0.2296$ (approximately) of $\beta_{e}$ such that the interval of $k$ for the existence of the curve described by $PQ=0$ in $k-n_{2}$ plane increases with increasing $\beta_{e}$ for $ 0 \leq \beta_{e} < \beta_{e}^{(c)} $. Also, from figure \ref{P4_PQ_eq_zero_k_vs_n1_diff_beta_e_subplot}, we can conclude that the stable region of modulated IA waves described by the inequality $PQ<0$ increases with increasing $\beta_{e}$ for $ 0 \leq \beta_{e} < \beta_{e}^{(c)}$.  From figure \ref{P4_PQ_eq_zero_k_vs_n1_diff_beta_e_subplot_111}, we see that the interval of $k$ for the existence of the curve defined by $PQ=0$ decreases with increasing $\beta_{e}$ and it is simple to check that the interval of $k$ for the existence of the curve obtained from the relation $PQ=0$ decreases with increasing $\beta_{e}$ for $\beta_{e}^{(c)} < \beta_{e} < \frac{4}{7}$. Also, from figure \ref{P4_PQ_eq_zero_k_vs_n1_diff_beta_e_subplot_111}, we see that both the regions $PQ<0$ and $PQ>0$  decreases with increasing $\beta_{e}$ for $\beta_{e}^{(c)} < \beta_{e} < \frac{4}{7}$. So, we can conclude that the stable region of modulated IA wave described by the inequality $PQ<0$ decreases with increasing $\beta_{e}$ for $\beta_{e}^{(c)} < \beta_{e} < \frac{4}{7}$. But for both the figures (figures \ref{P4_PQ_eq_zero_k_vs_n1_diff_beta_e_subplot} - \ref{P4_PQ_eq_zero_k_vs_n1_diff_beta_e_subplot_111}), for any point $(k,n_{2})$ lying within the region described by the inequality $PQ>0$, the modulated IA wave is stable or unstable according to whether $K \geq K_{c}$ or $K<K_{c}$.

In figure \ref{P4_PQ_eq_zero_k_vs_n1_diff_n_sc}, $n_{2}$ is plotted against $k$ when $PQ=0$ for different values of $n_{sc}$: (a) $n_{sc}=0.01$, (b) $n_{sc}=0.3$, (c) $n_{sc}=0.6$ and (d) $n_{sc}=0.9$  when $\gamma=5/3$, $\sigma=0.001$, $\sigma_{sc}=0.25$, $\omega_{c}=0.2$ and $\beta_{e}=0.1$. From figure \ref{P4_PQ_eq_zero_k_vs_n1_diff_n_sc}, we see that the curve in $k-n_{2}$ plane obtained from the relation $PQ=0$ is not a continuous curve, there exist at least two discontinuities. In figure \ref{P4_PQ_eq_zero_k_vs_n1_diff_n_sc}(a), we have seen that the curve $PQ = 0$ exist within the intervals $0.168 < k <	0.192$ and $1.138 < k <	1.818$. In figure \ref{P4_PQ_eq_zero_k_vs_n1_diff_n_sc}(b), we have seen that the curve $PQ = 0$ exist within the intervals $ 0.172 < k <	0.198$ and $0.718 < k < 1.59$. In figure \ref{P4_PQ_eq_zero_k_vs_n1_diff_n_sc}(c), we have seen that the curve $PQ = 0$ exist within the intervals $0.174 < k < 0.2$ and $0.84 < k <	1.694$. And finally in figure \ref{P4_PQ_eq_zero_k_vs_n1_diff_n_sc}(d), we have seen that the curve $PQ = 0$ exist within the intervals $0.174 < k <	0.2$ and $0.956 < k <	1.768$. Therefore, figure \ref{P4_PQ_eq_zero_k_vs_n1_diff_n_sc} shows that the modulated IA wave is stable for any point $(k, n_{2})$ lying within the region(s) described by the inequality $PQ<0$. Again, for any point $(k,n_{2})$ lying within the region defined by $PQ>0$, modulated IA wave is stable or unstable according to whether $K \geq K_{c}$ or $K<K_{c}$.

In figure \ref{P4_PQ_eq_zero_k_vs_n1_diff_omegha_c_subplot}, $n_{2}$ is plotted against $k$ when $PQ=0$ for different values of $\omega_{c}$: (a) $\omega_{c}=0.1$, (b) $\omega_{c}=0.2$, (c) $\omega_{c}=0.3$ and (d) $\omega_{c}=0.4$ when $\gamma=5/3$, $\sigma=0.001$, $\sigma_{sc}=0.25$, $n_{sc}=0.9$ and $\beta_{e}=0.1$. From this figure, we can conclude that the interval of $k$ for the existence of the curve defined by the relation $PQ=0$ in $k-n_{2}$ plane decreases with increasing $\omega_{c}$. We also observe that the stable region of modulated IA waves defined by the inequality $PQ<0$ increases with increasing $\omega_{c}$ whereas the region in $k-n_{2}$ plane described by $PQ>0$ decreases with increasing $\omega_{c}$ and for any point $(k,n_{2})$ lying within the region defined by $PQ>0$, modulated IA wave is stable or unstable according to whether $K \geq K_{c}$ or $K<K_{c}$.

Analytically we have studied the instability condition and the maximum growth rate of instability ($\Gamma_{max}$) in the section  \ref{P4_sec:Modulation_instability}. We have also investigated the maximum growth rate of instability ($\Gamma_{max}$) with the help of figures \ref{P4_Gamma_max_diff_beta_e}, \ref{P4_Gamma_max_diff_n_sc}, \ref{P4_Gamma_max_diff_omega_c} and \ref{P4_Gamma_max_diff_n_1} to show the effect of different parameters on $\Gamma_{max}$.

In figure \ref{P4_Gamma_max_diff_beta_e}, $P$, $Q$ and $\Gamma_{max}/|\phi_{0}|^{2}$ are plotted in \ref{P4_Gamma_max_diff_beta_e}(a), \ref{P4_Gamma_max_diff_beta_e}(b) and \ref{P4_Gamma_max_diff_beta_e}(c), respectively, against $k$ for different values of $\beta_{e}$ and for $\gamma=5/3$, $\sigma=0.001$, $n_{sc}=0.25$, $\sigma_{sc}=0.25$, $\omega_{c}=0.2$ and $n_{2}=0.25$. In figure \ref{P4_Gamma_max_diff_beta_e}(a) there is a zoomed region of the figure \ref{P4_Gamma_max_diff_beta_e}(a) within the rectangle: $0.6 \leq k \leq 2$, $0 \leq P \leq 0.0025$.  From figure \ref{P4_Gamma_max_diff_beta_e}, we see that the maximum growth rate of instability of the modulated IA wave increases with increasing $\beta_{e}$. There are two different types of curves: Type-I and Type-II. Type-I looks like green curve whereas red and blue curves are of Type-II. In fact, there exists a certain value $\beta_{e}^{(c)}=0.1286$ (approximately) of $\beta_{e}$ such that if $\beta_{e}<\beta_{e}^{(c)}$ then the curve is of Type-I and if $\beta_{e}>\beta_{e}^{(c)}$ then the curve is of Type-II. In this figure, we see that the green curve exists within the interval $ 0.816 < k \leq 2$ only and consequently, for $k$ lying within the interval $ 0 < k < 0.816 $, $PQ<0$, and consequently, modulated IA wave is stable. Again, for $k$ lying within the intervals $0 < k <0.08$ and $0.594 < k < 0.752$, the red curve does not exist because within these intervals of $k$ ($0 < k <0.08$ \& $0.594 < k < 0.752$), $PQ<0$, and consequently, modulated IA wave is stable. Also, for $k$ lying within the intervals $0<k<0.004$ and $0.546<k<0.69$, the blue curve does not exist because within these intervals of $k$ ($0<k<0.004$ \& $0.546<k<0.69$), $PQ<0$, i.e., the modulated IA wave is stable. From this figure, we see that there exists a critical value $k_{c}$ of $k$ such that for any fixed $k$ lying within the interval $0<k \leq k_{c}$, the maximum growth rate of instability of the modulated IA wave (if exists) increases with increasing values of $\beta_{e}$. Again, for any fixed $\beta_{e}$, the maximum growth rate of instability of the modulated IA wave (if exists) increases with increasing $k$.  

In figure \ref{P4_Gamma_max_diff_n_sc}, $P$, $Q$ and $\Gamma_{max}/|\phi_{0}|^{2}$ are plotted in \ref{P4_Gamma_max_diff_n_sc}(a), \ref{P4_Gamma_max_diff_n_sc}(b) and \ref{P4_Gamma_max_diff_n_sc}(c), respectively, against $k$ for different values of $n_{sc}$ with $\sigma=0.001$, $\beta_{e}=0.3$, $\sigma_{sc}=0.25$, $\omega_{c}=0.2$ and $n_{2}=0.25$. In figure \ref{P4_Gamma_max_diff_n_sc}(a) there is a zoomed region of the figure \ref{P4_Gamma_max_diff_n_sc}(a) within the rectangle: $0.67 \leq k \leq 2$ , $0 \leq P \leq 0.002$.  From this figure, we see that the green curve does not exists within the intervals $ 0 < k <  0.15$ and $ 0.468 < k <  0.688$ and consequently, for $k$ lying within the intervals $ 0 < k <  0.15$ and $ 0.468 < k <  0.688$, we have $PQ<0$, and consequently the modulated IA wave is stable. Again, for $k$ lying within the intervals $0 < k <0.08$ and $0.594 < k < 0.752$, the red curve does not exist because within these intervals of $k$, $PQ<0$, and consequently the modulated IA wave is stable. Also, for the blue curve, for $k$ lying within the intervals $0<k<0.494$ and $0.658<k<0.788$,  the modulated IA wave is stable. From this figure, we see that for any fixed $n_{sc}$, the maximum growth rate of instability of the modulated IA wave (if exists) increases with increasing $k$.

For $\beta_{e}=0$, i.e., when both the electron species are isothermally distributed then for this case, in figure \ref{P4_Gamma_max_diff_omega_c}, $P$, $Q$ and $\Gamma_{max}/|\phi_{0}|^{2}$ are plotted in (a), (b) and (c), respectively, against $k$ for different values of $\omega_{c}$. From this figure, we see that the region of existence of maximum growth rate of instability decreases with increasing $\omega_{c}$, i.e., the maximum growth rate of instability of the modulated wave decreases with increasing $\omega_{c}$. There exists a critical value $\omega_{c}^{(c)}=0.4595$ (approximately) of $\omega_{c}$ such that the maximum growth rate of instability disappears from the system for $\omega_{c}>\omega_{c}^{(c)}$. From this figure, we see that for any fixed $\omega_{c}$, the maximum growth rate of instability of the modulated IA wave (if exists) increases with increasing $k$ whereas for any fixed $k$, the maximum growth rate of instability of the modulated IA wave (if exists) decreases with increasing $\omega_{c}$. 

In figure \ref{P4_Gamma_max_diff_n_1}, $P$, $Q$ and $\Gamma_{max}/|\phi_{0}|^{2}$ are plotted in (a), (b) and (c), respectively, against $k$ for different values of $n_{2}$ with $\sigma=0.001$, $\beta_{e}=0.1$, $\sigma_{sc}=0.25$, $\omega_{c}=0.2$ and $n_{sc}=0.25$. From this figure \ref{P4_Gamma_max_diff_n_1}, we see that the green curve exists within the intervals $0.427 < k < 0.507$ and $0.662 < k \leq 2$ as $PQ>0$ for all values of $k$ lying within the interval $0.427 < k < 0.507$ or $0.662 < k \leq 2$. For all values of $k$ lying within the interval $0.507 < k < 0.662$, $ PQ < 0$, and consequently, the modulated IA wave is stable for $0.507 < k < 0.662$. From figure \ref{P4_Gamma_max_diff_n_1}, for any fixed value of $k$, the maximum growth rate of instability of modulated IA wave (if exists) increases with increasing $n_{2}$ whereas for any fixed value of $n_{2}$, the maximum growth rate of instability increases with increasing $k$.

\section{CONCLUSIONS}

In this paper, we have studied the MI of obliquely propagating IA waves in a magnetized collisionless plasma consisting of adiabatic warm ions and two species of electrons at different temperatures, a cooler one with a Boltzmann distribution and a hotter one with a nonthermal Cairns [9] distribution, immersed in a uniform static magnetic field directed along $z-$axis. 

We have made a systematic analysis to derive a NLSE by using the Reductive Perturbation Method (RPM) [34, 35]. We have seen that the dispersion relation of the IA wave propagating at an arbitrary angle to the direction of the magnetic field is exactly same as that of the IA wave propagating parallel to the direction of the magnetic field if $\theta=0$, where $\theta$ is the angle between the direction of propagation of the IA wave and the direction of the magnetic field. We have seen that the group velocity of the IA wave propagating at an arbitrary angle to the direction of the magnetic field is exactly same as that of the IA wave propagating parallel to the direction of the magnetic field if $\theta=0$.
Finally, it is possible to make a correspondence between the MI of IA wave propagating at an arbitrary angle to the direction of the magnetic field and the MI of IA wave propagating parallel to the direction of the magnetic field.      

It is observed that the region of existence of maximum growth rate of instability decreases with increasing values of the magnetic field intensity whereas the region of existence of the maximum growth rate of instability increases with increasing $\cos \theta$. Again, the maximum growth rate of instability increases with increasing $\cos \theta$ and also this maximum growth rate of instability increases with increasing $\beta_{e}$ upto a critical value of the wave number, where $\beta_{e}$ is the parameter associated with the nonthermal distribution of hotter electron species.

\textbf{REFERENCES}

1. P. O. Dovner, A. I. Eriksson, R. Bostr{\"o}m, and B. Holback, Geophys. Res. Lett. \textbf{21}, 1827 (1994).

2. R. Bostr{\"o}m, G. Gustafsson, B. Holback, G. Holmgren, H. Koskinen, and P. Kintner, Phys. Rev. Lett. \textbf{61}, 82 (1988).

3. R. Bostr{\"o}m, IEEE Trans. Plasma Sci. \textbf{20}, 756 (1992).

4. R. E. Ergun, C. W. Carlson, J. P. McFadden, F. S. Mozer, G. T. Delory, W. Peria, C. C.
Chaston, M. Temerin, R. Elphic, R. Strangeway, et al., Geophys. Res. Lett. \textbf{25}, 2025
(1998).

5. R. E. Ergun, C. W. Carlson, J. P. McFadden, F. S. Mozer, G. T. Delory, W. Peria, C. C.
Chaston, M. Temerin, R. Elphic, R. Strangeway, et al., Geophys. Res. Lett. \textbf{25}, 2061
(1998).

6. G. T. Delory, R. E. Ergun, C. W. Carlson, L. Muschietti, C. C. Chaston, W. Peria, J. P.
McFadden, and R. Strangeway, Geophys. Res. Lett. \textbf{25}, 2069 (1998).

7. R. Pottelette, R. E. Ergun, R. A. Treumann, M. Berthomier, C. W. Carlson, J. P. McFadden, and I. Roth, Geophys. Res. Lett. \textbf{26}, 2629 (1999).

8. J. P. McFadden, C. W. Carlson, R. E. Ergun, F. S. Mozer, L. Muschietti, I. Roth, and
E. Moebius, J. Geophys. Res. \textbf{108}, 8018 (2003).

9. R. A. Cairns, A. A. Mamum, R. Bingham, R. Bostr{\"o}m, R. O. Dendy, C. M. C. Nairn,
and P. K. Shukla, Geophys. Res. Lett. \textbf{22}, 2709 (1995).

10. M. Y. Yu and H. Luo, Phys. Plasmas \textbf{15}, 024504 (2008).

11. H. R. Pakzad and M. Tribeche, Astrophys. Space Sci. \textbf{334}, 45 (2011).

12. H. G. Abdelwahed, Astrophys. Space Sci. \textbf{341}, 491 (2012).

13. M. S. Alam, M. M. Masud, and A. A. Mamun, Chin. Phys. B \textbf{22}, 115202 (2013).

14. S. S. Ghosh and A. N. Sekar Iyengar, Phys. Plasmas \textbf{21}, 082104 (2014).

15. I. Tasnim, M. M. Masud, and A. A. Mamun, Plasma Phys. Rep. \textbf{40}, 723 (2014).

16. S. V. Singh and G. S. Lakhina, Commun. Nonlinear Sci. Numer. Simul. \textbf{23}, 274 (2015).
 
17. Y. W. Hou, M. X. Chen, M. Y. Yu, and B. Wu, Plasma Phys. Rep. \textbf{42}, 900 (2016). 

18. M. A. Hossen and A. A. Mamun, IEEE Trans. Plasma Sci. \textbf{44}, 643 (2016).

19. D. N. Gao, J. Zhang, Y. Yang, and W. S. Duan, Plasma Phys. Rep. \textbf{43}, 833 (2017). 

20. S. Islam, A. Bandyopadhyay, and K. P. Das, J. Plasma Phys. \textbf{74}, 765 (2008).

21. O. R. Rufai, R. Bharuthram, S. V. Singh, and G. S. Lakhina, Phys. Plasmas \textbf{21}, 082304 (2014).

22. S. Dalui, A. Bandyopadhyay, and K. P. Das, Phys. Plasmas \textbf{24}, 042305 (2017).

23. S. Dalui, A. Bandyopadhyay, and K. P. Das, Phys. Plasmas \textbf{24}, 102310 (2017).

24. R. Sabry, W. Moslem, and P. Shukla, Plasma Phys. Control. Fusion \textbf{54}, 035010 (2012).
 
25. S. A. El-Tantawy, A. M. Wazwaz, and A. Rahman, Phys. Plasmas \textbf{24}, 022126 (2017). 

26. E. Parkes, J. Phys. A: Math. Gen. \textbf{20}, 3653 (1987).

27. M. K. Alam and A. R. Chowdhury, Aust. J. Phys. \textbf{53}, 289 (2000).

28. A. P. Misra and C. Bhowmik, Phys. Plasmas \textbf{14}, 012309 (2007).

29. A. Bains, A. Misra, N. Saini, and T. Gill, Phys. Plasmas \textbf{17}, 012103 (2010).

30. G. Murtaza and M. Salahuddin, Plasma Phys. \textbf{24}, 451 (1982).

31. N. Jehan, M. Salahuddin, H. Saleem, and A. M. Mirza, Phys. Plasmas \textbf{15}, 092301 (2008).
 
32. T. Kakutani and N. Sugimoto, Phys. Fluids \textbf{17}, 1617 (1974).

33. B. Ghosh and S. Banerjee, Turk. J. Phys. \textbf{40}, 1 (2016).

34. T. Taniuti and N. Yajima, J. Math. Phys. \textbf{10}, 1369 (1969).

35. N. Asano, T. Taniuti, and N. Yajima, J. Math. Phys. \textbf{10}, 2020 (1969).


\end{document}